\documentclass[aps,prd,nofootinbib]{revtex4-2}
\usepackage{amsmath,amssymb}
\usepackage{graphicx}
\usepackage{dsfont}

\newcommand{\be}{\begin{equation}}
\newcommand{\ee}{\end{equation}}
\newcommand{\ba}{\begin{eqnarray}}
\newcommand{\ea}{\end{eqnarray}}
\newcommand{\ban}{\begin{eqnarray*}}
\newcommand{\ean}{\end{eqnarray*}}
\newcommand \nn {\nonumber}

\begin{document}

\title{Stability of Classical Chromodynamic Fields -- Addendum}

\author{Sylwia Bazak$^1$\footnote{e-mail: sylwia.bazak@gmail.com} and Stanis\l aw Mr\' owczy\' nski$^{1,2}$\footnote{e-mail: stanislaw.mrowczynski@ncbj.gov.pl}}

\affiliation{$^1$Institute of Physics, Jan Kochanowski University, ul. Uniwersytecka 7, PL-25-406 Kielce, Poland 
\\
$^2$National Centre for Nuclear Research, ul. Pasteura 7,  PL-02-093 Warsaw, Poland}

\date{August 18, 2022}

\begin{abstract}

A system of chromodynamic fields, which can be treated as classical, is generated at the earliest stage of relativistic heavy-ion collisions. Numerical simulations show that the system is unstable but the nature of the instability is not well understood. We study the problem systematically. In the first paper, we have performed a linear stability analysis of space-time uniform chromoelectric and chromomagnetic fields. There they have been considered the Abelian configurations of single-color potentials linearly depending on coordinates and nonAbelian ones where the fields are generated by the multi-color non-commuting uniform potentials. Here we extend and supplement the analysis. We discuss the parallel chromoelectric and chromomagnetic fields which occur simultaneously. We also consider a general nonAbelian configurations of the uniform fields. Finally, we discuss the gauge dependence of our results. 

\end{abstract}

\maketitle

\section{Introduction}

Numerical simulations of the earliest phase of relativistic heavy-ion collisions performed in the framework of the Color Glass Condensate (CGC) approach -- reviewed at length in the articles \cite{Iancu:2003xm,Gelis:2012ri} -- show that the system is unstable \cite{Romatschke:2005pm,Romatschke:2006nk}, see also \cite{Fukushima:2007yk}. However, as explained in detail in \cite{Bazak:2021xay}, a character of the instability is not well understood. We plan to study the problem systematically. Since the system under consideration is described in terms of classical fields in the CGC approach used in \cite{Romatschke:2005pm,Romatschke:2006nk}, we analyze a stability of classical chromodynamic fields. 

In the first part of our project \cite{Bazak:2021xay}, we studied a linear stability of constant and uniform chromomagnetic and chromoelectric fields that occur separately. We considered the Abelian configurations discussed in the past where the fields are due to single-color potentials linearly depending on coordinates. However, we were mostly focused on the nonAbelian configurations where the fields are generated by the multi-color non-commuting constant uniform potentials. In contrast to the Abelian configurations that satisfy the sourceless Yang-Mills equations, the nonAbelian configurations are the solutions of Yang-Mills equations with appropriately chosen four-currents. We derived a complete spectrum of eigenfrequencies of small fluctuations around the background fields that obey the linearized Yang-Mills equations. The spectra of Abelian and nonAbelian configurations are similar but different and they both include unstable modes. We also briefly discussed the relevance of our results for fields that are uniform only in a limited spatial domain.

Here we discuss a system of parallel chromoelectric and chromomagnetic fields as according to the CGC approach such a configuration occurs at the earliest phase of relativistic heavy-ion collisions. In Sec.~\ref{sec-stability} we show that the system of such fields, which are constant and uniform, can be generated only by the potential known from the Abelian theory. The nonAbelian configuration is not possible. We perform the stability analysis showing that the system's dynamics is dominated by the chromoelectric field. The solutions of linearized Yang-Mills equations run away either to plus or minus infinity and in this sense the system is genuinely unstable. 

As already mentioned, we studied in \cite{Bazak:2021xay} a stability of nonAbelian configurations of space-time uniform  chromoelectric and chromomagnetic fields. Here we extend the analysis, considering in Sec.~\ref{sec-nonAbelian} a whole class of nonAbelian configurations that yield the same field strengths and the same energy density but are still gauge inequivalent, that is, they cannot be changed one into another by means of a gauge transformation \cite{Brown:1979bv}. It has been recently  conjectured \cite{Vachaspati:2022ktr}, see also \cite{Pereira:2022lbl}, that such configurations of chromoelectric fields are stable against the quantum Schwinger process of spontaneous production of gluons. However, our linear analysis shows that the configurations of both chromoelectric and chromomagnetic fields are classically unstable. 

Since we use the linearized Yang-Mills equations and we fix a gauge condition, one wonders how our results depend on a chosen gauge. In Sec.~\ref{sec-EM-tensor} we show that, in spite of the linearization the chromodynamic strength tensor transforms covariantly under gauge transformations, and consequently quantities like an energy-momentum tensor, which are obtained from the strength tensor, are gauge invariant. We compute the energy momentum tensor corresponding to solutions of the linearized Yang-Mills equations and we demonstrate that in case of unstable modes the system's energy density and other components of the energy momentum tensor exponentially grow in time. 

For completeness of our presentation, we introduce in Sec.~\ref{sec-linear-QCD} the linearized Yang-Mills equations in the background gauge, which were already discussed in \cite{Bazak:2021xay}. The equations are subsequently used in the stability analyses. We close our study in Sec.~\ref{sec-conclusions}, summarizing and concluding our results.

Throughout the paper the indices $i,j = x, y, z$ and $\mu, \nu = 0, 1, 2, 3$ label, respectively, the Cartesian spatial coordinates and those of Minkowski space. The signature of the metric tensor is $(+,-,-,-)$.  The indices $a, b = 1, 2, \dots N_c^2 -1$ numerate color components in the adjoint representation of the SU($N_c$) gauge group. However, we mostly use the SU(2) group. We neglect henceforth the prefix `chromo' when referring to chromoelectric or chromomagnetic fields. Since we study chromodynamics only, this should not be confusing.

\section{Linearized Classical Chromodynamics}
\label{sec-linear-QCD}

The Yang-Mills equations written in the adjoint representation of  the SU($N_c$) gauge group are
\be
\label{YM-eqs}
D^{ab}_\mu F_b^{\mu \nu} = j_a^\nu, 
\ee
where $D^{ab}_\mu \equiv \partial_\mu \delta^{ab} - g f^{abc}A^c_\mu$, $j_a^\nu$ is the color current and the strength tensor is 
\be
\label{F1}
F^{\mu\nu}_a=\partial^\mu A^\nu_a - \partial^\nu A^\mu_a + g f_{abc} A^\mu_b A^\nu_c .
\ee
The electric and magnetic fields are given as
\be
\label{E-B-fields-def}
E_a^i = F_a^{i0} , ~~~~~~~~~~
B_a^i = \frac{1}{2} \epsilon^{ijk} F_a^{kj} ,
\ee
where $\epsilon^{ijk}$ is the Levi-Civita fully antisymmetric tensor.

We assume that the potential $\bar{A}_a^\mu$ solves the Yang-Mills equation (\ref{YM-eqs}) and we consider small fluctuations $a_a^\mu$ around $\bar{A}_a^\mu$. So, we define the potential 
\be
\label{A=Abar+a}
A_a^\mu(t, {\bf r}) \equiv \bar{A}_a^\mu(t, {\bf r})  + a_a^\mu(t, {\bf r}),
\ee
such that $|\bar{A}(t, {\bf r})| \gg |a(t, {\bf r})|$.

Assuming that the background potential $\bar{A}^\mu_a$ satisfies the Lorenz gauge condition $\partial_\mu \bar{A}^\mu_a = 0$ while the fluctuation potential $a_a^\mu$ that of the background gauge 
\be
\label{background-gauge}
\bar{D}^{ab}_\mu a_b^\mu = 0,
\ee
where $\bar{D}^{ab}_\mu \equiv \partial_\mu \delta^{ab} - g f^{abc}\bar{A}^c_\mu$, the Yang-Mills equation linearized in $a_a^\mu$ can be written as
\be
\label{linear-YM-background}
\big[ g^{\mu \nu}(\bar{D}_\rho \bar{D}^\rho)_{ac} + 2 g f^{abc} \bar{F}_b^{\mu\nu} \big] a^c_\nu = 0 .
\ee

The background gauge appears particularly convenient for our purposes because different color and space-time components of $a^a_\mu$ are mixed only through the tensor $\bar{F}_b^{\mu\nu}$ which enters Eq.~(\ref{linear-YM-background}). In case of other gauges, {\it e.g.}, the Lorenz gauge $\partial_\mu a^\mu_a = 0$, the mixing is more severe. Throughout the stability analysis presented in \cite{Bazak:2021xay} and here, we use the background gauge that facilitates comparisons of various cases.

\section{Parallel electric and magnetic fields}
\label{sec-stability}

The electric and magnetic fields that are constant uniform and both along the axis $x$ are generated by the potential known from the Abelian theory, that is
\be
\label{pot-abel}
\bar{A}_a^\mu(t, {\bf r}) = \delta^{a1}(-xE,0,0,yB)
\ee
where ${\bf r}=(x,y,z)$. Our stability analysis is limited to the SU(2) gauge group when $ f^{abc} = \epsilon^{abc}$ with $a,b = 1, 2, 3$. Using Eqs.~(\ref{E-B-fields-def}), one finds that  the only non-vanishing elements of the tensor $\bar{F}_a^{\mu \nu}$ corresponding to the potential (\ref{pot-abel}) are $\bar{F}_1^{x0}=-\bar{F}_1^{0x} = E$ and $\bar{F}_1^{zy} = - \bar{F}_1^{yz} = B$ that is
\be
\label{E-B-fields}
{\bf E}_a(t, {\bf r}) = \delta^{a1}(E,0,0) 
~~~~~~~~~~
{\bf B}_a(t, {\bf r}) = \delta^{a1}(B,0,0).
\ee
The potential (\ref{pot-abel}) solves the equations of motion (\ref{YM-eqs}) with vanishing current. The nonAbelian terms disappear because there is only one color component. The potential satisfies the Lorenz gauge condition $\partial_\mu \bar{A}^\mu_a = 0$.

We note that the electric and magnetic fields (\ref{E-B-fields}) are of the same color. If the potential is chosen, for example, as $\bar{A}_a^\mu(t, {\bf r}) = \delta^{a1}(-xE,0,0,0) +\delta^{a2}(0,0,0,yB)$, then except the uniform electric and magnetic fields of the colors 1 and 2, respectively, there is the non-uniform electric field along the axis $z$ of color 3 corresponding to $\bar{F}_3^{0z} = - g x y E B = - \bar{F}_3^{z0}$. 

When the electric and magnetic fields are studied separately, the uniform electric and magnetic fields can occur not only due to potentials linearly depending on coordinates as in the Abelian theory, but also due to non-linear terms of the strength tensor (\ref{F1}) \cite{Bazak:2021xay}. However, there is no analogous nonAbelian configuration that generates only parallel and uniform electric and magnetic fields. The reason is as follows. The nonAbelian contributions to the electric and magnetic fields in the SU(2) Yang-Mills theory are $E^i_a = g \epsilon^{abc}\bar{A}^i_b \bar{A}^0_c$ and $B^i_a = g \epsilon^{ijk}\epsilon^{abc} \bar{A}^j_b \bar{A}^k_c$. To get the electric and magnetic fields along the axis $x$ there must be non-vanishing components of $\bar{A}^0_a,~\bar{A}^x_a$ and $\bar{A}^y_a,~\bar{A}^z_a$. The potentials $\bar{A}^0_a$ and $\bar{A}^x_a$ must be of different colors and the same holds for $\bar{A}^y_a$ and $\bar{A}^z_a$. If we choose $\bar{A}^0_2, ~\bar{A}^x_3$ and $\bar{A}^y_2,~\bar{A}^z_3$ we get the nonzero electric and magnetic fields $E^x_1$ and $B^x_1$ but additionally there appear the fields $E^z_1$ and $B^z_1$. If we choose $\bar{A}^0_2, ~\bar{A}^x_3$ and $\bar{A}^y_1,~\bar{A}^z_2$ there are except the fields $E^x_1$ and $B^x_3$ also $E^y_3$, $B^z_2$, and $B^y_1$. So, one observes that there is no way to have the parallel electric and magnetic fields and no other fields. We conclude that the potential (\ref{pot-abel}) is the unique configuration with the constant and uniform electric and magnetic fields along the axis $x$.

Let us now consider a stability of the configuration (\ref{pot-abel}). The linearized Yang-Mills equation (\ref{linear-YM-background}) is
\be
\label{a-mu-bcg}
\Box a^\mu_a 
+ 2g \epsilon^{ab1}(xE \partial_0 -  By \partial_z ) a^\mu_b 
- 2g \epsilon^{ab1} (E\delta^{\mu 0} a^x_b + E\delta^{\mu x} a^0_b  
+ B \delta^{\mu y} a^z_b - B \delta^{\mu z} a^y_b)
+ g^2 (E^2 x^2 - B^2 y^2) \epsilon^{ac1} \epsilon^{cb1} a^\mu_b = 0 .
\ee
Defining the functions
\be
\label{T-X-Y-Z-def}
T^\pm = a^0_2 \pm i a^0_3 , ~~~~~~~~
X^\pm = a^x_2 \pm i a^x_3 , ~~~~~~~~
Y^\pm = a^y_2 \pm i a^y_3 , ~~~~~~~~
Z^\pm = a^z_2 \pm i a^z_3 , 
\ee
Eq.~(\ref{a-mu-bcg}) provides
\ba
\label{Tplus-bcg}
&&\Box T^+ 
- 2i g (xE \partial_0 -  By \partial_z ) T^+ 
+ 2i g E X^+  
- g^2 (E^2 x^2 - B^2 y^2) T^+ = 0 ,
\\[2mm]
\label{Xplus-bcg}
&&\Box X^+ 
- 2ig (xE \partial_0 -  By \partial_z ) X^+ 
+ 2ig E T^+ 
- g^2 (E^2 x^2 - B^2 y^2) X^+ = 0 ,
\\[2mm]
\label{Tminus-bcg}
&&\Box T^- 
+ 2ig (xE \partial_0 -  By \partial_z ) T^- 
- 2ig E X^- 
- g^2 (E^2 x^2 - B^2 y^2) T^- = 0 ,
\\[2mm]
\label{Xminus-bcg}
&&\Box X^- 
+ 2ig (xE \partial_0 -  By \partial_z ) X^- 
- 2i g E T^- 
- g^2 (E^2 x^2 - B^2 y^2) X^- = 0 ,
\\[2mm]
\label{Yplus-bcg}
&&\Box Y^+ 
- 2ig (xE \partial_0 -  By \partial_z ) Y^+ 
+ 2i g B Z^+ 
- g^2 (E^2 x^2 - B^2 y^2) Y^+ = 0 ,
\\[2mm]
\label{Zplus-bcg}
&&\Box Z^+ 
- 2ig (xE \partial_0 -  By \partial_z ) Z^+ 
- 2ig B Y^+
- g^2 (E^2 x^2 - B^2 y^2) Z^+ = 0 ,
\\[2mm]
\label{Yminus-bcg}
&&\Box Y^- 
+ 2ig (xE \partial_0 -  By \partial_z ) Y^- 
- 2ig B Z^- 
- g^2 (E^2 x^2 - B^2 y^2) Y^- = 0 ,
\\[2mm]
\label{Zminus-bcg}
&&\Box Z^- 
+ 2ig (xE \partial_0 -  By \partial_z ) Z^- 
+ 2ig B Y^-
- g^2 (E^2 x^2 - B^2 y^2) Z^- = 0 .
\ea

To diagonalize the equations (\ref{Tplus-bcg}) - (\ref{Zminus-bcg}) we introduce the functions
\be
\label{G-H-U-W-def}
G^\pm \equiv T^+ \pm X^+,
~~~~~~~~~~~
H^\pm \equiv T^- \pm X^-,
~~~~~~~~~~~
U^\pm \equiv Y^+ \pm iZ^+ ,
~~~~~~~~~~~
W^\pm \equiv Y^- \pm iZ^- ,
\ee
and assuming that the functions $a^\mu_a$ depend on $t, z$ as $e^{-i(\omega t - k_z z)}$, the equations (\ref{Tplus-bcg}) - (\ref{Zminus-bcg}) yield
\ba
\label{Gplus-bcg-k2}
&&\Big(- \frac{\partial^2}{\partial x^2} - \frac{\partial^2}{\partial y^2}
 -(\omega + gEx)^2 + (k_z- g B y)^2
+ 2i g E  \Big)G^+ = 0 ,
\\[2mm]
\label{Gminus-bcg-k2}
&&\Big(- \frac{\partial^2}{\partial x^2} - \frac{\partial^2}{\partial y^2} 
 -(\omega + gEx)^2 + (k_z- g B y)^2
- 2ig E  \Big)G^- = 0 ,
\\[2mm]
\label{Hplus-bcg-k2}
&&\Big(- \frac{\partial^2}{\partial x^2} - \frac{\partial^2}{\partial y^2}  
 -(\omega - gEx)^2 + (k_z + g B y)^2
- 2ig E \Big) H^+ = 0 ,
\\[2mm]
\label{Hminus-bcg-k2}
&&\Big(- \frac{\partial^2}{\partial x^2} - \frac{\partial^2}{\partial y^2}  
 -(\omega - gEx)^2 + (k_z + g B y)^2
+ 2i g E \Big) H^- = 0 ,
\\[2mm]
\label{Uplus-bcg-k2}
&&\Big( - \frac{\partial^2}{\partial x^2} - \frac{\partial^2}{\partial y^2} 
 -(\omega + gEx)^2 + (k_z- g B y)^2
+ 2 g B \Big) U^+ = 0 ,
\ea
\ba
\label{Uminus-bcg-k2}
&&\Big(- \frac{\partial^2}{\partial x^2} - \frac{\partial^2}{\partial y^2}  
 -(\omega + gEx)^2 + (k_z- g B y)^2
- 2g B  \Big) U^- = 0 ,
\\[2mm]
\label{Wplus-bcg-k2}
&&\Big( - \frac{\partial^2}{\partial x^2} - \frac{\partial^2}{\partial y^2}  
 -(\omega - gEx)^2 + (k_z + g B y)^2
- 2g B  \Big) W^+ = 0 ,
\\[2mm]
\label{Wminus-bcg-k2}
&&\Big(- \frac{\partial^2}{\partial x^2} - \frac{\partial^2}{\partial y^2}  
 -(\omega - gEx)^2 + (k_z + g B y)^2
+ 2g B  \Big) W^- = 0 .
\ea

The equations (\ref{Gplus-bcg-k2}) - (\ref{Wminus-bcg-k2}) can be solved by the variable separation method. We solve, for example, the equations (\ref{Uplus-bcg-k2}) and (\ref{Uminus-bcg-k2}). Assuming that 
\be
\label{separation-U}
U^{\pm} (x,y) = U^{\pm}_E (x) \, U^{\pm}_B (y) ,
\ee
one finds two equations
\ba
\label{UE-eq}
\Big[C^\pm_U - \frac{\partial^2}{\partial x^2} 
- g^2 E^2 \Big(\frac{\omega}{gE} + x \Big)^2 \Big] U^\pm_E(x)  &=& 0 ,
\\[1mm]
\label{UB-eq}
\Big[- C^\pm_U  \pm 2 g B - \frac{\partial^2}{\partial y^2} 
+ g^2B^2 \Big(\frac{k_z}{g B} -  y\Big)^2 \Big] U^\pm_B(y) &=&  0 ,
\ea
where $C^\pm_U$  is the separation constant. 

Under the replacements 
\be
C^\pm_U  \mp 2 g B ~ \rightarrow ~2m{\cal E},  ~~~~~~~~~~ 
g B ~ \rightarrow ~ m \bar\omega, ~~~~~~~~~~
\frac{k_z}{gB}  ~ \rightarrow ~ y_0 ,
\ee
where ${\cal E}$ is the energy of harmonic oscillator and $\bar\omega$ is the frequency of its classical counterpart, 
the equation (\ref{UB-eq}) coincides with the Schr\"odinger equation of the harmonic oscillator which can be written as
\be
\label{Schr-eq-HO}
\Big(- 2m{\cal E} + m^2 \bar\omega^2(y_0 - y )^2 - \frac{d^2}{dy^2} \Big) \varphi(y) = 0 .
\ee
Since the oscillator energy is quantized as
\be
{\cal E}_n = \bar\omega  \Big(n + \frac{1}{2}\Big), ~~~~~~ n = 0,\,1,\,2,\, \dots 
\ee
the separation constant equals
\be
C^\pm_U  = g B  (2 n + 1)  \pm 2 g B ,
\ee
and the solutions are well-known to be
\be
\label{UB-solution}
U^\pm_B(y)  \sim H_n\big(a(y-y_0)\big)\; e^{-\frac{1}{2}a^2 (y-y_0)^2} ,
\ee
where  $a \equiv \sqrt{g B}$, $y_0 \equiv \frac{k_z}{gB}$ and $H_n$ is the Hermite polynomial.

Under the replacements 
\be
C^\pm_U ~ \rightarrow ~ - 2m{\cal E},  ~~~~~~~~~~ 
g E ~ \rightarrow ~ m \bar\omega, ~~~~~~~~~~
\frac{\omega}{gE}  ~ \rightarrow ~ x_0 ,
\ee
the equation (\ref{UE-eq}) coincides with the Schr\"odinger equation of the inverted harmonic oscillator which can be written as
\be
\label{Schr-eq-IHO}
\Big(- 2m{\cal E} - m^2 \bar\omega^2(x_0 - x )^2 - \frac{d^2}{dx^2} \Big) \phi(x) = 0 .
\ee

As discussed in detail in \cite{Barton:1984ey}, there are no normalizable solutions of the Schr\"odinger equation of the inverted harmonic oscillator. The solutions run away either to plus or minus infinity and thus the configuration of the constant electric field is genuinely unstable. 

\section{NonAbelian Configurations of Uniform Fields}
\label{sec-nonAbelian}

\subsection{Magnetic Field}
\label{sec-nonAbelian-B}

A nonAbelian configuration of $\bar{A}^\mu_a$ that produces a constant homogeneous magnetic field ${\bf B}_a = \delta^{a1}(B,0,0)$ is
\ba
\label{Abar-matrix}
\bar{A}^\mu_a = 
\left[ {\begin{array}{cccc}
0 ~~~~& 0 & 0 & 0 \\
0 ~~~~& 0 & 0 & \lambda \sqrt{B/g} \\
0 ~~~~& 0 & \frac{1}{\lambda}\sqrt{B/g} & 0 \\
 \end{array} } \right]  ,
\ea
where the Lorenz index $\mu$ numerates the columns and the color index $a$ numerates the rows; $\lambda$ is an arbitrary real number different from zero. In the first paper \cite{Bazak:2021xay}, we studied in detail the special case of the potential (\ref{Abar-matrix}) with $\lambda =1$. Here we briefly discuss the general case of arbitrary $\lambda$.

The potential (\ref{Abar-matrix}), which obviously satisfies the Lorenz gauge condition, does not solve the Yang-Mills equation (\ref{YM-eqs}) with $j_a^\mu=0$. Computing $\bar{D}^{ab}_\mu \bar{F}_b^{\mu \nu}$, one finds
\ba
\label{eq-j-B-matrix}
\left[ {\begin{array}{cccc}
0 ~~~~& 0 & 0 & 0 \\
0 ~~~~& 0 & 0 &  \frac{1}{\lambda} g^{1/2} B^{3/2} \\
0 ~~~~& 0 & \lambda g^{1/2} B^{3/2} & 0 \\
 \end{array} } \right] = j^\mu_a .
\ea
Following \cite{Tudron:1980gq}, we assume that the current, which enters the Yang-Mills equation, equals the left-hand side of Eq.~(\ref{eq-j-B-matrix}). Then, the potential (\ref{Abar-matrix}) solves the Yang-Mills equations (\ref{YM-eqs}). 

Since the current $j^\mu$ in the fundamental representation transforms under a gauge transformation $U$ as $j^\mu \to Uj^\mu U^\dagger$ the quantity ${\rm Tr}[j^\mu j_\mu]$ is a gauge invariant and so is $j^\mu_a j_{a \,\mu}$. The current (\ref{eq-j-B-matrix}) yields
\be
\label{jj}
j^\mu_a j_{a \,\mu} = - \Big(\frac{1}{\lambda^2} +\lambda^2\Big) g B^3 .
\ee
Because the gauge invariant (\ref{jj}) depends on $(\lambda^{-2}+\lambda^2)$ the potential configurations (\ref{Abar-matrix}) of different $(\lambda^{-2}+\lambda^2)$ are gauge inequivalent \cite{Brown:1979bv}, even so the configurations produce the same field strength and the same energy density, see Sec.~\ref{sec-EM-tensor}. Therefore, it is of physical interest to analyze the stability of the configuration (\ref{Abar-matrix}) of arbitrary $\lambda$.  

The equation of motion of the small field $a^\mu_a$ (\ref{linear-YM-background}) is found to be
\be
\label{a-mu-bcg-nA}
\Box a^\mu_a 
+ 2g A (\lambda^{-1}\epsilon^{a3b}\partial_y + \lambda \epsilon^{a2b}\partial_z ) a^\mu_b
- g^2 A^2 (\lambda^2 \epsilon^{a2e}\epsilon^{e2b} 
+ \lambda^{-2} \epsilon^{a3e}\epsilon^{e3b} ) a^\mu_b
+ 2g^2 A^2 \epsilon^{a1b}( \delta^{\mu y} a^z_b -  \delta^{\mu z} a^y_b) 
 = 0 ,
\ee
where $A \equiv \sqrt{B/g}$.

Assuming that $a_a^\mu (t,x,y,z) = e^{-i (\omega t - {\bf k}\cdot {\bf r})} a_a^\mu$, where ${\bf k} = (k_x, k_y, k_z)$ and  ${\bf r} = (x,y,z)$, Eqs.~(\ref{a-mu-bcg-nA}) are changed into the following set of algebraic equations
\be
\label{eqs-B}
\hat{M}_B^t \,\vec{a^t} = 0,
~~~~~~~~~~~
\hat{M}_B^x \,\vec{a^x} = 0,
~~~~~~~~~~~
\hat{M}_B^{yz} \,\vec{a^{yz}} = 0,
\ee
where $\vec{a^t}$ and $\vec{a^x}$ represent three colors of the corresponding components of the four-vector $a^\mu$. The vector $\vec{a^{yz}}$ is six dimensional; it includes three colors of $a^y$ and $a^z$. The matrices $\hat{M}_B^t$ and $\hat{M}_B^x$, which are equal to each other, are $3 \times 3$ and $\hat{M}_B^{yz}$ is $6 \times 6$. The explicit form of the matrices for $\lambda =1$ is given in \cite{Bazak:2021xay}.

The dispersion equations read 
\be
\label{dis-eqs-B}
{\rm det}\hat{M}_B^t  = 0,
~~~~~~~~~~~
{\rm det}\hat{M}_B^x  = 0,
~~~~~~~~~~~
{\rm det}\hat{M}_B^{yz} = 0.
\ee
In \cite{Bazak:2021xay} we have found a complete set of analytical solutions of  Eqs.~(\ref{dis-eqs-B}) for $\lambda =1$. The solutions of the first two equations are always stable, while there is an unstable solution (that is with positive imaginary part) of the third equation. Since the analysis of the general case of $\lambda \not=1$ is more complex it is limited here to ${\bf k} = 0$ which is sufficient to show that the system is unstable. 

One finds three solutions of the first two identical equations (\ref{dis-eqs-B}) which are
\be
\omega_0^2 = \lambda^2 gB, 
~~~~~~~~~
\omega_+^2 = \frac{1 + \lambda^4}{\lambda^2} \, gB, 
~~~~~~~~~
\omega_-^2 =  \frac{1}{\lambda^2} \, gB .
\ee
Since $\omega_0^2$, $\omega_+^2$ and $\omega_-^2$ are real and positive, similar to the special case $\lambda =1$, there are all stable modes corresponding to these solutions.
 
The solutions of the third equation (\ref{dis-eqs-B}) read
\be
\omega_1^2 = \frac{1 + \lambda^4}{\lambda^2} \, gB, 
~~~~~~~~~
\omega_2^2 = \frac{1 + \lambda^4 + \sqrt{1 + 14 \lambda^4 + \lambda^8}}{2 \lambda^2} \, gB, 
~~~~~~~~~
\omega_3^2 = \frac{1 + \lambda^4 - \sqrt{1 + 14 \lambda^4 + \lambda^8}}{2 \lambda^2} \, gB .
\ee
One checks that $\omega_3^2 < 0$ for $\lambda^2 > 0$. Therefore, as in the special case $\lambda = 1$, there is the unstable mode associated with the overdamped mode. 

\subsection{Electric Field}
\label{sec-nonAbelian-E}

A nonAbelian configuration of $\bar{A}^\mu_a$, which produces a constant homogeneous electric field along the axis $x$ is 
\ba
\label{pot-nonabel}
\bar{A}_a^\mu = 
\left[ {\begin{array}{cccc}
0 & 0 & 0 & 0 \\
\lambda \sqrt{E/g} & 0 & 0 & 0 \\
0 &  \frac{1}{\lambda}\sqrt{E/g} & 0 & 0 \\
  \end{array} } \right] ,
\ea
where, as previously, $\lambda$ is an arbitrary real number different from zero. One checks that the potential (\ref{pot-nonabel}) yields the electric field ${\bf E}_a = \delta^{a1}(-E,0,0)$. In our first paper \cite{Bazak:2021xay}, where we studied in detail the special case of the potential (\ref{pot-nonabel}) with $\lambda =1$, we incorrectly wrote that the field is of opposite direction that is ${\bf E}_a = \delta^{a1}(E,0,0)$.  Since the error does not influence our analysis in any way, we also use the potential (\ref{pot-nonabel}) here to simplify a comparison of the results and to avoid confusion. However, we discuss the general case of arbitrary $\lambda$.

The potential (\ref{pot-nonabel}), which obviously satisfies the Lorenz gauge condition, does not solve the Yang-Mills equation (\ref{YM-eqs}) with $j_a^\mu=0$. Instead, one gets
\ba
\label{eq-j-E-matrix}
\left[ {\begin{array}{cccc}
0 & 0 & 0 & 0 \\
\frac{1}{\lambda}g^{1/2} E^{3/2} & 0 & 0 & 0 \\
0 & - \lambda g^{1/2} E^{3/2} & 0 & 0 \\
 \end{array} } \right] = j^\mu_a .
\ea
As in the case of magnetic field we assume that the current, which enters the Yang-Mills equation, equals the left-hand side of Eq.~(\ref{eq-j-E-matrix}).  Then, the potential (\ref{pot-nonabel}) solves the Yang-Mills equations (\ref{YM-eqs}). 

The current (\ref{eq-j-B-matrix}) provides the gauge invariant 
\be
\label{jj-E}
j^\mu_a j_{a \,\mu} = \Big(\frac{1}{\lambda^2} - \lambda^2\Big) g E^3 ,
\ee
and thus the potential configurations (\ref{pot-nonabel}) of different $(\lambda^{-2}-\lambda^2)$ are gauge inequivalent \cite{Brown:1979bv}; even so, the configurations produce the same field strength and the same energy density. 

The equation of motion of the small field $a^\mu_a$ (\ref{linear-YM-background}) is found to be
\be
\label{a-mu-nA}
\Box a^\mu_a 
+ 2g A (\lambda \epsilon^{a2b}\partial_0 + \lambda^{-1}\epsilon^{a3b}\partial_x ) a^\mu_b
+ g^2 A^2 (\lambda^2 \epsilon^{a2e}\epsilon^{e2b} - \lambda^{-2}\epsilon^{a3e}\epsilon^{e3b} ) a^\mu_b
+ 2g^2 A^2 \epsilon^{a1b}( \delta^{\mu 0} a^x_b +  \delta^{\mu x} a^0_b) = 0 ,
\ee
where $A \equiv \sqrt{E/g}$.

Assuming that $a_a^\mu (t,x,y,z) = e^{-i (\omega t - {\bf k}\cdot {\bf r})} a_a^\mu$, where ${\bf k} = (k_x, k_y, k_z)$ and  ${\bf r} = (x,y,z)$, Eqs.~(\ref{a-mu-nA}) are changed into the following set of algebraic equations
\be
\label{eqs-E}
\hat{M}_E^{tx} \,\vec{a^{tx}} = 0,
~~~~~~~~~~
\hat{M}_E^y \,\vec{a^y} = 0,
~~~~~~~~~~
\hat{M}_E^z \,\vec{a^z} = 0. 
\ee
The second and third equation are identical to each other.  In \cite{Bazak:2021xay} we have found a complete set of analytical solutions of  the dispersion equations
\be
\label{dis-eqs-E}
{\rm det}\hat{M}_E^{tx} = 0,
~~~~~~~~~~
{\rm det} \hat{M}_E^y = 0,
~~~~~~~~~~
{\rm det} \hat{M}_E^z = 0  
\ee
for $\lambda =1$. There are unstable solutions of all three equations. The analysis of the general case of $\lambda \not=1$ is limited here to ${\bf k} = 0$, but it still reveals that the system is unstable. 

The three solutions of the second and third identical equations (\ref{dis-eqs-E}) are
\be
\omega_1^2 = \frac{1}{2}\Big(2\lambda^2 + \frac{1}{\lambda^2} + \sqrt{8 + \frac{1}{\lambda^4}}\, \Big) gE,
~~~~~~~~~
\omega_2^2 = \frac{1}{\lambda^2} \, gE, 
~~~~~~~~~
\omega_3^2 = \frac{1}{2}\Big(2\lambda^2 + \frac{1}{\lambda^2} - \sqrt{8 + \frac{1}{\lambda^4}}\, \Big) gE. 
\ee
As one observes $\omega_3^2 < 0$ for $\lambda^2 < 1$ and then there is the unstable mode. 

The first equation (\ref{dis-eqs-E}) is effectively cubic in $\omega^2$. There are one real and two complex solutions which can be found using the Cardano formula. The solutions are
\be
\omega_1^2 = u + v  - \frac{1}{3}a_2,
~~~~~~~~
\omega_2^2 = -\frac{1}{2}(u + v) + \frac{i \sqrt{3}}{2}(u - v) - \frac{1}{3}a_2,
~~~~~~~~
\omega_3^2 = -\frac{1}{2}(u + v) - \frac{i \sqrt{3}}{2}(u - v) - \frac{1}{3}a_2,
\ee
with
\be
u \equiv \sqrt[3]{-\frac{q}{2} + \sqrt{\frac{q^2}{4} + \frac{p^3}{27}}} ,
~~~~~~~~~~
v \equiv \sqrt[3]{-\frac{q}{2} - \sqrt{\frac{q^2}{4} + \frac{p^3}{27}}} ,
\ee
and
\be
p \equiv -\frac{1 - 7\lambda^4 + \lambda^8}{3\lambda^4}\, g^2 E^2 ,
~~~~~~~~
q \equiv \frac{2 - 21\lambda^4 + 141\lambda^8 + 2 \lambda^{12}}{27\lambda^6}\, g^3 E^3 ,
~~~~~~~~
a_2  \equiv - \frac{2(1 + \lambda^4)}{\lambda^2}\, g E.
\ee
One checks that $\omega_1^2 < 0$ for $\lambda^2 >1$ and then there is the unstable mode. Since the imaginary parts of $\omega_2^2$ and $\omega_3^2$ are nonzero there are also unstable modes related to these solutions. So, we conclude that the configuration of electric field generated by the potential (\ref{pot-nonabel}) is severely unstable.

\section{Energy-Momentum Tensor}
\label{sec-EM-tensor}

Solutions of the Yang-Mills equations do not have a direct physical meaning because of their dependence on a chosen gauge. However, there are gauge invariant quantities that are determined by the solutions. The energy-momentum tensor of chromodynamic fields is a particularly important example as it can tell us how system's energy density and other tensors' components change in time. In this way we can observe, for example, whether the system evolves toward thermodynamic equilibrium. 

As discussed in the review article \cite{Blaschke:2016ohs}, the energy-momentum tensor of pure Yang-Mills theory, which is gauge invariant, divergenceless, symmetric, and traceless, is
\be
\label{EMT}
T^{\mu\nu} = 2{\rm Tr}\big[F^{\mu\rho}F_\rho^{~\nu} 
+\frac{1}{4}g^{\mu\nu}F^{\sigma\tau}F_{\sigma\tau}\big] 
= F_a^{\mu\rho}F_{\rho\,a}^{~\nu}+\frac{1}{4}g^{\mu\nu}F_a^{\sigma\tau}F_{\sigma\tau\,a}.
\ee
The expressions after the first and second equality are written in the fundamental and adjoint representation, respectively. Using the formulas (\ref{E-B-fields-def}), the elements of the energy-momentum tensor (\ref{EMT}) can be expressed through the electric and magnetic fields. We are particularly interested in the diagonal elements: energy density $\varepsilon$ and pressures $p_L$ and $p_T$, which are given as
\ba
\label{epsilon}
\varepsilon \equiv T^{00} &=& \frac{1}{2} \big({\bf E}_a \cdot {\bf E}_a + {\bf B}_a \cdot {\bf B}_a \big) ,
\\[1mm]
\label{pL}
p_L \equiv T^{xx} &=& -E_a^x E_a^x - B_a^x B_a^x + \varepsilon ,
\\[1mm]
\label{pT}
p_T \equiv T^{yy} &=& -E_a^y E_a^y - B_a^y B_a^y + \varepsilon .
\ea
Since we consider the background of electric and magnetic fields along the axis $x$, the subscripts $L$ and $T$ refer the this direction.  

\subsection{Gauge Dependence}

The energy-momentum tensor (\ref{EMT}) is gauge invariant because the strength tensor in the fundamental representation transforms under the gauge transformation $U$ as 
\be
\label{gauge-trans-F}
F^{\mu \nu} \rightarrow U F^{\mu \nu} U^\dagger .
\ee 
One wonders whether the tensor (\ref{EMT}) is still gauge invariant within the linearized chromodynamics where non-linear terms, which contribute to the strength tensor $F^{\mu \nu}$, are neglected. 

The gauge potential, which is the sum of the background and fluctuation potentials, transforms as 
\be
\label{gauge-trans-Abar+a}
A^\mu =  \bar{A}^\mu  + a^\mu \rightarrow 
U \bar{A}^\mu U^\dagger + 
U a^\mu U^\dagger + 
\frac{i}{g} U \partial^\mu U^\dagger, 
\ee
but there is an ambiguity how to transform $\bar{A}^\mu$ and $a^\mu$ separately. However, the background potential $\bar{A}^\mu$ should behave as the gauge potential independent of the fluctuation potential $a^\mu$ which, in particular, can vanish. Therefore, the background potential $\bar{A}^\mu$ transforms as a gauge potential, which in turn dictates the transformation of the fluctuation potential $a^\mu$. Consequently, the transformation is
\be
\label{gauge-trans-Abar-a}
\bar{A}^\mu \rightarrow U\bar{A}^\mu U^\dagger  
+\frac{i}{g} U \partial^\mu U^\dagger  ,
~~~~~~~~~~~~~~~~~~
a^\mu \rightarrow U a^\mu U^\dagger  .
\ee
We note that the transformation (\ref{gauge-trans-Abar-a}) is used in the background field method, see {\it e.g.} \cite{Abbott:1981ke}.

One observes that the strength tensor corresponding to $A^\mu =  \bar{A}^\mu  + a^\mu$, which equals
\be
F^{\mu \nu} = \bar{F}^{\mu \nu} + \partial^\mu a^\nu - \partial^\nu a^\mu
 - ig [a^\mu, a^\nu]  - ig [\bar{A}^\mu, a^\nu]  - ig [a^\mu, \bar{A}^\nu] ,
\ee
transforms according to Eq.~(\ref{gauge-trans-F}) even when the term quadratic in $a^\mu$ is neglected. The point is that the term $[a^\mu, a^\nu]$ transforms as $[a^\mu, a^\nu] \rightarrow U[a^\mu, a^\nu]U^\dagger$. Consequently, the quantity ${\rm Tr}\big[F^{\mu\lambda}F^{\rho\sigma}\big]$ and, in particular, the energy-momentum tensor  (\ref{EMT}) are gauge invariant under the transformation (\ref{gauge-trans-Abar-a}).

\subsection{Energy density and pressures of stable and unstable modes}

We consider here, as an example, the energy density (\ref{epsilon}) and pressures (\ref{pL}), (\ref{pT}) corresponding to the two modes of the potential fluctuating around the Abelian configuration of uniform and constant magnetic field $B$ along the axis $x$ which were studied in \cite{Bazak:2021xay} and denoted as $U^\pm$. The mode $U^+$ is always stable, while the mode $U^-$ is the well-known Nielsen-Olesen instability \cite{Nielsen:1978rm} for  $k_x^2 < gB$ and $n=0$. Otherwise, the mode  $U^-$ is stable.

The modes are found in terms of the functions $U^\pm$ but the energy-momentum tensor (\ref{EMT}) is expressed through the gauge potential. So, we have to reconstruct the potential. Keeping in mind the definitions (\ref{T-X-Y-Z-def}) and (\ref{G-H-U-W-def}), one finds 
\ba
U^+ = a^y_2 + i a^y_3 +i a^z_2 - a^z_3,
\\[1mm]
U^- = a^y_2 + i a^y_3 -i a^z_2 + a^z_3.
\ea
Demanding that $U^- = W^\pm = 0$ for the mode $U^+$ and that $U^+ = W^\pm = 0$ for the mode $U^-$, one finds the gauge potentials
\ba
\label{A-matrix}
A^\mu_a = 
\left[ {\begin{array}{cccc}
0 ~&~ 0 & 0 & yB \\
0 ~&~ 0 & f & -if \\
0 ~&~ 0 & -i f & -f \\
 \end{array} } \right]  ,
~~~~~~~~~~~~~~~~~~
A^\mu_a = 
\left[ {\begin{array}{cccc}
0 ~&~ 0 & 0 & yB \\
0 ~&~ 0 & f & if \\
0 ~&~ 0 & -i f & f \\
 \end{array} } \right]  ,
\ea
which correspond to the mode $U^+$ or $U^-$. The function $f$ equals $h(y) e^{-i(\omega t - k_x x - k_z z)}$ for the stable mode $U^+$ and $h(y) e^{\gamma t}e^{i(k_x x + k_z z)}$ for unstable mode $U^-$. The parameters $\omega, \gamma$ are real. The function $h(y)$ will be defined below. We note that the columns in Eq.~(\ref{A-matrix}) correspond to Cartesian components of the fields while the rows to the color components. 

Keeping in mind that only real parts of the potentials (\ref{A-matrix}) have a physical meaning, the electric and magnetic fields associated with the $U^+$ and $U^-$ modes are, respectively 
\ba
\label{E-B-U+}
{\bf E}_a
&=& \left[ {\begin{array}{ccc}
   0 ~& 0  & 0  \\
   0 ~&~  -  \partial^t f^R ~&~ -\partial^t f^I  \\
   0 ~&~ -\partial^t f^I ~&~  \partial^t f^R \\
  \end{array} } \right] , ~~~~~~~~~~~~~~
{\bf B}_a
= \left[ {\begin{array}{ccc}
    B & 0  & 0  \\
    - \partial^y f^I + \partial^z f^R - g y B f^I ~&~  \partial^x f^I ~&~  -\partial^x f^R  \\
   \partial^y f^R + \partial^z f^I + g y B f^R ~&~ -\partial^x f^R ~&~ -\partial^x f^I \\
  \end{array} } \right] ,
\\[5mm]
\label{E-B-U-}
{\bf E}_a
&=& \left[ {\begin{array}{ccc}
   0 ~& 0  & 0  \\
   0 ~&~  -  \partial^t f^R ~&~ \partial^t f^I  \\
   0 ~&~ -\partial^t f^I ~&~  -\partial^t f^R \\
  \end{array} } \right] , ~~~~~~~~~~~~~~
{\bf B}_a
= \left[ {\begin{array}{ccc}
    B & 0  & 0  \\
   \partial^y f^I + \partial^z f^R - g y B f^I ~&~  \partial^x f^I ~&~  -\partial^x f^R  \\
   -\partial^y f^R + \partial^z f^I + g y B f^R ~&~ \partial^x f^R ~&~ -\partial^x f^I \\
  \end{array} } \right] ,
\ea
where $f^R \equiv \Re f$ and $f^I \equiv \Im f$. The terms quadratic in $f$ are neglected in Eqs.~(\ref{E-B-U+}) and (\ref{E-B-U-}). 

The energy density, longitudinal and transverse pressures associated with the mode $U^+$ are
\ba
\nn
\varepsilon &=& \frac{1}{2} B^2 + (\partial^t f^R)^2 + (\partial^t f^I)^2 + (\partial^x f^I)^2 + (\partial^x f^R)^2
+  \frac{1}{2} (- \partial^y f^I + \partial^z f^R - g y B f^I)^2 
\\ \label{e+}
&&~~~~~~~~~~~~~~~~~~~~~~~~~~~~~~~~~
~~~~~~~~~~~~~~~~~~~~~~~~~~~~\,
+ \frac{1}{2} (\partial^y f^R - \partial^z f^I + g y B f^R)^2 ,
\\[2mm] \label{pL+}
p_L  &=&  - B^2 - (- \partial^y f^I + \partial^z f^R - g y B f^I)^2 
- (\partial^y f^R + \partial^z f^I + g y B f^R)^2  + \varepsilon ,
\\[2mm] \label{pT+}
p_T  &=& - (\partial^t f^R)^2 - (\partial^t f^I)^2  - (\partial^x f^I)^2  - (\partial^x f^R)^2 + \varepsilon ,
\ea
and those of the mode $U^-$ read
\ba
\nn
\varepsilon   &=& \frac{1}{2} B^2  + (\partial^t f^R)^2 + (\partial^t f^I)^2 + (\partial^x f^I)^2 +  (\partial^x f^R)^2
+ \frac{1}{2} (\partial^y f^I + \partial^z f^R - g y B f^I)^2 
\\ \label{e-}
&&~~~~~~~~~~~~~~~~~~~~~~~~~~~~~~~~\,
~~~~~~~~~~~~~~~~~~~~~~~~~~~~~
+ \frac{1}{2} (-\partial^y f^R + \partial^z f^I + g y B f^R)^2 
\\[2mm] \label{pL-}
p_L  &=& - B^2 - (\partial^y f^I + \partial^z f^R - g y B f^I)^2 
- ( -\partial^y f^R + \partial^z f^I + g y B f^R)^2 + \varepsilon ,
\\[2mm] \label{pT-}
p_T  &=&  - (\partial^t f^R)^2 - (\partial^t f^I)^2 - (\partial^x f^I)^2 - (\partial^x f^R)^2 + \varepsilon .
\ea
We note that all the terms, which enter the electric and magnetic fields (\ref{E-B-U+}) and (\ref{E-B-U-}), are kept here to have gauge invariant expressions. 

Further on, we consider the lowest energy modes $U^\pm$ when $h(y) = \delta e^{-\frac{g B y^2}{2} }$ with $\delta$ being the fluctuation's amplitude of the dimension of mass which is small that is $\delta \ll \sqrt{B}$. For simplicity, we also put $k_x = k_z = 0$ and then, $\gamma^2 = gB$, $\omega^2 = 3gB$ \cite{Bazak:2021xay}. The energy density, longitudinal and transverse pressures of the stable mode are 
\ba
\label{e+final}
\varepsilon &=& \frac{1}{2} B^2 + gB \delta^2 (3 + 2 g y^2B) e^{- g B y^2}  , 
\\[1mm]
\label{pL+final}
p_L &=&  -  \frac{1}{2} B^2 + gB \delta^2 (3 - 2 g y^2B) e^{- g B y^2} ,
\\[1mm] 
\label{pT+final}
p_T &=&  \frac{1}{2} B^2 + 2 g^2y^2 B^2 \delta^2 e^{- g B y^2},
\ea
and those of the unstable mode equal 
\ba
\label{e+final}
\varepsilon &=& \frac{1}{2} B^2  + gB \delta^2 e^{- g B y^2} e^{2\sqrt{gB} \, t} ,
\\[1mm]
\label{pL+final}
p_L &=& - \frac{1}{2}B^2 + gB \delta^2 e^{- g B y^2} e^{2\sqrt{gB} \, t} ,
\\[1mm] 
\label{pT+final}
p_T &=&   \frac{1}{2} B^2 .
\ea
We see that $\varepsilon$, $p_L$ and $p_T$ are time independent for the stable mode but exponentially grow in the case of instability. We also observe that the growing unstable mode tends to reduce the negative value of $p_L$, which can be interpreted as a beginning of the evolution toward equilibrium. 

The expressions (\ref{e+final}) -- (\ref{pT+final}) should be treated with caution. One checks that the biggest terms, which are neglected, are of the order $gB \delta^2$. Consequently, they are as big as the terms that are kept in $\varepsilon$, $p_L$, and $p_T$. If we consistently neglect the terms of the order $gB \delta^2$, the energy density and longitudinal pressure of the stable and unstable modes are determined solely by the background potential. We note that the term $gB y^2 e^{-gBy^2}$ is always smaller or equal to $1/e$.

\section{Summary, conclusions and outlook}
\label{sec-conclusions}

We have discussed the system of classical electric and magnetic fields that are parallel to each other, constant and uniform. Such a system can be generated only by the single-color potential linearly depending on coordinates which is well known in the Abelian theory. In contrast to the electric or magnetic fields, which occur separately, there is no nonAbelian configuration where the fields are generated by the multi-color non-commuting constant uniform potentials. We have performed a linear stability analysis of the system of parallel electric and magnetic fields, showing that the system's dynamics is dominated by the electric field. The solutions of linearized Yang-Mills equations run away either to plus or minus infinity, which makes the system unstable. 

There is a whole class of nonAbelian potentials of both electric and magnetic fields that correspond to the same field strength and energy density but are still gauge inequivalent. We have shown that all these field configurations are unstable in the long wavelength limit. 

We have discussed the gauge dependence of our results and we have shown that in spite of the linearization the chromodynamic strength tensor transforms covariantly under gauge transformations. Therefore, the energy-momentum tensor, which is obtained from the strength tensor, is gauge invariant. We have demonstrated that the tensor corresponding to the unstable Nielsen-Olesen mode exponentially grows in time starting an evolution toward equilibrium. 

With the present paper we close our discussion of constant and uniform chromodynamic fields. The next steps are field configurations that are more relevant for relativistic heavy-ion collisions. We are going to perform a stability analysis of the background fields, which are invariant under Lorenz boosts in one direction, with fluctuations that break the Lorenz invariance. This is the configuration where the instability was found in \cite{Romatschke:2005pm,Romatschke:2006nk}. 

\section*{Acknowledgments}

We are grateful to Giorgio Torrieri for inspiring criticism and to Jerzy Kowalski-Glikman and Margaret Carrington for useful discussions. This work was partially supported by the National Science Centre, Poland under grant 2018/29/B/ST2/00646. 


\end{document}